%% file: main.tex
\documentclass[conference]{IEEEtran}
\IEEEoverridecommandlockouts
\usepackage{cite,comment}
\usepackage{amsmath,amssymb,amsfonts}
\usepackage{algorithmic}
\usepackage{algorithm}
\usepackage{graphicx}
\usepackage{textcomp}
\usepackage{xcolor}
\usepackage{braket}
\usepackage{float}
\usepackage{booktabs}
\usepackage{subcaption}
\usepackage{balance}

\def\BibTeX{{\rm B\kern-.05em{\sc i\kern-.025em b}\kern-.08em
    T\kern-.1667em\lower.7ex\hbox{E}\kern-.125emX}}
\begin{document}

 
\title{Distributing Arbitrary Quantum Graph States by Graph Transformation\\
}
\author{Tingxiang Ji,
        Jianqing Liu,
        Zheshen Zhang
\thanks{
T. Ji and J. Liu are with the Department
of Computer Science, North Carolina State University, Raleigh,
NC 27606 USA, e-mail: \{tji2, jliu96\}@ncsu.edu}
\thanks{Z. Zhang is with the Department of Electrical Engineering and Computer Science, University of Michigan, Ann Arbor, MI 48109 USA, e-mail: zszh@umich.edu}
\thanks{The work by J. Liu was supported in part by the National Science Foundation under grants OMA-2304118 and OMA-2326746. The work by Z. Zhang was supported in part by EECS-2326780.}
}

\maketitle

\begin{abstract}
Quantum graph state is a special class of nonlocal state among multiple quantum particles, underpinning several nonclassical and promising applications such as quantum computing and quantum secret sharing. Recently, establishing quantum graph states among physically distant nodes has gained increasing popularity owing to its potential in expanding current quantum applications in scale. Existing research on this topic relies on a two-step approach: first distributing low-dimension elementary entanglement to target nodes, and then fusing them into a high-dimension quantum graph state. However, most existing studies focus solely on minimizing costs (e.g., the number of elementary entanglements consumed) to entangle target nodes, while neglecting the structure of the final quantum graph state. This can easily result in weak system entanglement, jeopardizing the graph state under partial measurement or noises. 

In this paper, we aim to establish any arbitrary quantum graph states of strong entanglement structures at a much lower cost than the state of the art. The method is to search for and establish an alternative state to the target state that is of lowest cost in creation. Subsequently, we transform such an alternative state back to the target state via compressed single-qubit Clifford operations.
To verify the performance of our developed algorithm, we conduct comprehensive simulations based on an open dataset containing all graph state structures up to 8 qubits. The results demonstrate fast algorithm convergence, an increased success probability in distributing any graph states, and $53.57\%$ saving in ERP cost compared with the state-of-the-art baseline. 

\end{abstract}


\input{introduction}
\input{preliminaries}
\input{system}

\input{principle}
\input{evaluation}

\input{conclusion}

\balance

\bibliographystyle{IEEEtran}
\bibliography{info}

\input{app}


\end{document}

%% file: introduction.tex
\section{Introduction}\label{introduction}

Significant progress in physical sciences and engineering has facilitated the development of operational quantum computers \cite{arute2019quantum, simonov2023universal}. However, the capabilities of individual quantum computers remain significantly limited, necessitating a paradigm shift towards interconnected systems\cite{li2021building,yu2022topology}. By linking smaller quantum units through quantum networks, we can forge extensive and resilient quantum computing infrastructures, thereby enabling distributed quantum computing and overcoming the limited capacity of single quantum computers \cite{cacciapuoti2019quantum}. Such networks are crucial for various critical applications, ranging from quantum key distribution for ultra-secure communications to quantum sensor networks for precision measurements surpassing classical limits\cite{djordjevic2024entanglement, azuma2023quantum, wehner2018quantum,ekert1991quantum}. 


The efficiency in generating and distributing entangled states over long distances serves as the foundational pillar for quantum networks and the associated quantum applications. This crucial aspect predominantly revolves around basic bipartite EPR pairs \cite{perseguers2008entanglement, schoute2016shortcuts,pirandola2016capacities} or more advanced multipartite entanglement, facilitating complex and interconnected applications \cite{pirker2018modular, hahn2019quantum, meignant2019distributing, fischer2021distributing,ghaderibaneh2023generation, chen2023multipartite,nguyen2022multiple}. Specifically, the multipartite entanglement is a nonlocal state shared among multiple particles, widely known as the quantum graph state. The notable N-partite GHZ state $\ket{\text{N-GHZ}}=\frac{1}{\sqrt{2}}(\ket{0}^{\otimes N}+\ket{1}^{\otimes N})$ is a special class of N-partite quantum graph state, equivalent through local transformations with $\frac{1}{\sqrt{2}}(\ket{0}\ket{+}^{\otimes N-1}+\ket{1}\ket{-}^{\otimes N-1})$. This equivalence illustrates the GHZ state as a star graph with one root and $N-1$ leaves. Researchers frequently utilize GHZ states to construct other multipartite entanglement states.

For instance, Pirker \emph{et al.} utilize the shared multipartite entanglement of the GHZ state in various sizes as a modular unit for constructing fully connected decorated graph states \cite{pirker2018modular}. Meignant \emph{et al.} achieve multipartite entanglement by initially identifying a Steiner tree that connects all selected physical network nodes, followed by employing star expansion techniques to generate the GHZ state with the least number of maximally entangled pairs derived from the Steiner tree. They further apply iterative star expansions to generate the Edge-Decorated Complete Graph (EDCG) state and employ Y-measurement and Z-measurement to attain any arbitrary graph state \cite{meignant2019distributing}. Fischer \emph{et al.} build upon the EDCG algorithm to construct graph states at a single node and then distribute the qubits to their designated nodes within the network, thereby achieving multipartite entanglement and establishing the upper bound on the elementary EPR pair consumption \cite{fischer2021distributing}. Ghaderibaneh \emph{et al.} have devised two distinct strategies, each comprising an initial phase of identifying the optimal node connections for distributing the GHZ state, followed by establishing a sequence of fusion operations on the entangled pairs generated across these connections \cite{ghaderibaneh2023generation}. Koudia et al. proposed a local operation based on state isometry from measurement and post-selection to create an arbitrary quantum graph state \cite{koudia2023quantum}.



These approaches are adept at establishing multipartite entanglement among selected nodes, primarily utilizing the GHZ state to forge entanglements. Yet, they often overlook the nuances of the specific target graph state and the quality of entanglement generated. 
First, GHZ state itself is more vulnerable to entanglement loss under the projective measurement on part of the system than other quantum graph states. 
Second, the graph states that existing works seek to establish, albeit the cost (e.g., EPR consumption) associated with their generation and distribution could be low, may not possess desired entanglement property for specific quantum applications. For example, a GHZ state may suffice for QKD or multi-party quantum conference key agreement protocols \cite{murta2020quantum}. Yet, one has to establish a 2D cluster state instead of a GHZ state for measurement-based quantum computing (MBQC), for it has been proven to be universal for MBQC \cite{van2006universal}. Among these applications, a high-fidelity entanglement is desired as a weak or poorly structured entanglement can undermine the effectiveness. To make this concrete, we analyze a set of 4-qubit graph states and compare their entanglement strength using the Schmidt measure \cite{eisert2001schmidt}.
For instance, consider the 4-qubit graph states depicted in Fig. \ref{fig:orbits}. When it comes to quantifying their degree of entanglement based on the Schdmit measure \cite{eisert2001schmidt}, the graph states in Fig. \ref{fig:orbits}(a), even the fully-connected graph state, are all equivalent to a GHZ state under local transformations, all of which have a Schmidt measure of 1; whereas the graph states in Fig. \ref{fig:orbits}(b) boast a Schmidt measure of 2, showing improvement in entanglement quality.

Moreover, many studies neglect the root position when utilizing GHZ states for multipartite entanglement distribution. That is, even among GHZ states with the same star topology, assigning different qubits (or network nodes) as the root can lead to different costs for GHZ state distribution in the underlying physical fiber network, thus affecting the overall cost of establishing the target multipartite entanglement. Nevertheless, the impact of GHZ root selection on distributing the GHZ state is often overlooked in existing research.


In this study, we explore the establishment of arbitrary graph states among remote spin qubits in cavities, mediated by photons flying over a noisy and lossy fiber network. We utilize the noise model from Muralidharan et al. \cite{muralidharan2014ultrafast}, which is well-suited to our specific cavity QED platform. The objective of this paper is then to investigate a graph state construction strategy that minimizes photonic EPR pair consumption while maximizing success probability. Instead of creating the target graph state directly, our idea is to search for and generate its local complementation (LC) equivalent state that results in a much higher creation success probability while consuming much less EPR pairs. Following this, we transform the LC-equivalent state back to the target graph state by applying only local operations. The realization of this idea follows a two-step approach. First, we develop a simulated annealing algorithm to efficiently identify the optimal LC-equivalent state within a prohibitively large search space for establishing elementary point-to-point EPR pairs. Next, we apply the quantum circuit compression technique to minimize the number of applied local operations for recovering the target graph state. To summarize, the key contributions of this paper are outlined as follows:
\begin{enumerate}
\item Our adapted simulated annealing algorithm demonstrates fast convergence within 5 iterations for graph states of up to 8 qubits and physical networks of up to 12 nodes, and meanwhile succeeds in identifying the globally optimal LC-equivalent state when the number of qubits is less than 6.
\item Based on a realistic qubit decoherence and loss model, it is calculated that our proposed approach can on average garner 4 orders of magnitude gain compared with directly distributing the target graph state.
\item Compared to the state-of-the-art techniques, our approach significantly reduces EPR consumption by up to $53.57\%$ for constructing any arbitrary graph state of up to 8 qubits. 
\end{enumerate}





The structure of this paper is organized as follows: Section \ref{preliminaries} provides an overview of the quantum preliminaries necessary for understanding the subsequent discussions. Section \ref{system} presents the quantum system model and assumptions. Section \ref{principle} delineates the working principles of our algorithm. Performance evaluation is covered in Section \ref{evaluation} followed by the discussion of key takeaways and open problems in Section \ref{discussion}. The paper is concluded in Section \ref{conclusion}, which also discusses future work. 


%% file: preliminaries.tex
\begin{figure*}[htbp] 
\centering 
\begin{subfigure}{0.40\textwidth}
  \centering
  \includegraphics[width=0.98\linewidth]{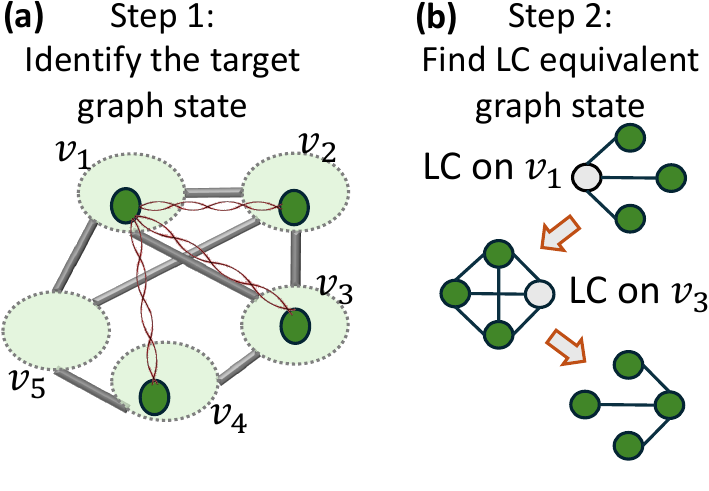}
\end{subfigure}
\hfill 
\begin{subfigure}{0.56\textwidth}
  \centering
  \includegraphics[width=\linewidth]{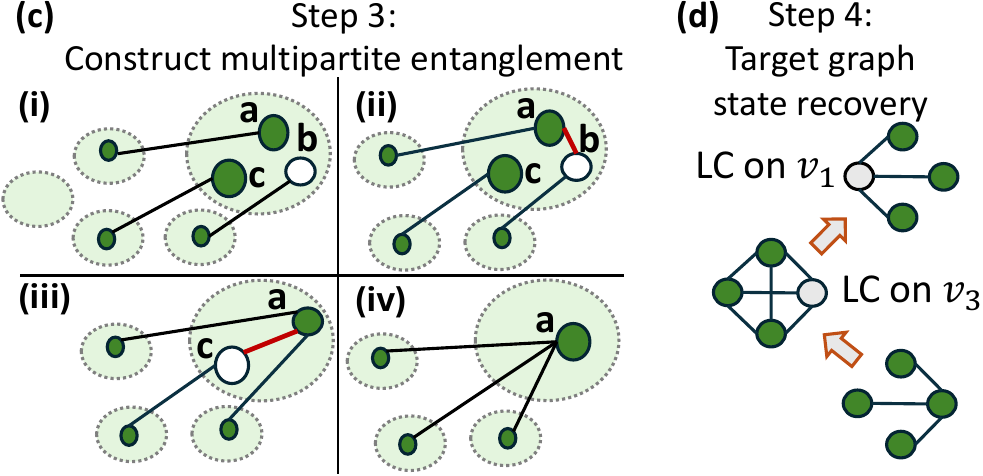}
\end{subfigure}

\caption{An example of distributing a 4-qubit graph state over a fiber-connected physical network. (a) Step 1 entails the application-specified entanglement structure (dark green nodes linked by wavy lines) and the physical network nodes (light green circles connected by solid black lines) holding qubits. (b) In Step 2, an LC-equivalent graph state is algorithmically identified that can offer a higher probability of success in entanglement distribution. (c) Step 3 constructs the LC-equivalent graph state via EPR distribution followed by qubit fusion. (d) Step 4 recovers the original graph state using LC operations.} 
\label{fig:framework}
\end{figure*}

\section{Preliminaries}\label{preliminaries}


To facilitate the understanding of our algorithm, we first introduce some preliminaries about qubit operators and quantum graph states.

\subsection{Qubit Operators}
\subsubsection{Pauli operators}
Pauli gates are single-qubit operators, comprising the pauli-$X$, $Y$, and $Z$ gates, defined as follows:
\begin{align*}
X &= \begin{pmatrix}
       0 & 1 \\
       1 & 0
     \end{pmatrix}, 
     & Y &= \begin{pmatrix}
              0 & -i \\
              i & 0
            \end{pmatrix}, 
            & Z &= \begin{pmatrix}
                     1 & 0 \\
                     0 & -1
                   \end{pmatrix} 
\end{align*}

The pauli-$X$ gate acts as a quantum bit-flip, mapping the basis state $\ket{0}$ to $\ket{1}$, and vice versa, thus $X\ket{0}=\ket{1}$, $X\ket{1}=\ket{0}$. 
The pauli-$Y$ gate gives rise to $Y\ket{0}=i\ket{1}$, $Y\ket{1}=-i\ket{0}$.
The pauli-$Z$ gate operates as a phase-flip, leaving $\ket{0}$ unchanged while adding a phase of $\pi$ to $\ket{1}$ state, hence $Z\ket{0}=\ket{0}$, $Z\ket{1}=-\ket{1}$. 

Another important single-qubit gate used in this paper is the Hadamard gate, defined as
$H =\frac{1}{\sqrt{2}} \begin{pmatrix}
        1 & 1 \\
        1 & -1
    \end{pmatrix}$.
This gate transforms $\ket{0}$ to $\ket{+}$, i.e., $H\ket{0}=\ket{+}=\frac{1}{\sqrt{2}}(\ket{0}+\ket{1})$, and $\ket{1}$ to $\ket{-}$, i.e., $H\ket{1}=\ket{-}=\frac{1}{\sqrt{2}}(\ket{0}-\ket{1})$.

\subsubsection{Controlled-$Z$ ($CZ$) Operation} 

The $CZ$ gate stands as a fundamental two-qubit operation that induces controlled phase shifts, thereby creating entanglement.

Mathematically, its matrix representation in the standard computational basis $\ket{00}$, $\ket{01}$, $\ket{10}$, $\ket{11}$ is defined as follows:

\begin{equation*}
CZ = 
\begin{pmatrix}
1 & 0 & 0 & 0 \\
0 & 1 & 0 & 0 \\
0 & 0 & 1 & 0 \\
0 & 0 & 0 & -1
\end{pmatrix}
\end{equation*}

This means that if the control qubit (the first qubit) is in state $\ket{0}$, the target qubit (the second qubit) is unchanged, i.e., $CZ\ket{00}=\ket{00}$, $CZ\ket{01}=\ket{01}$. If the control qubit is in state $\ket{1}$, a phase flip pauli-$Z$ gate is applied to the target qubit, i.e., $CZ\ket{10}=\ket{10}$, $CZ\ket{11}=-\ket{11}$.











\subsection{Density Matrix}

In quantum mechanics, the state of a system is often described using a density matrix, denoted as $\rho$. For a pure state $\psi$, $\rho$ is defined as $\rho=\ket{\psi}\bra{\psi}$, where $\ket{\psi}$ is state vector in the Hilbert space. For mixed states, which represent a statistical ensemble of pure states, $\rho=\sum_ip_i\ket{\psi_i}\bra{\psi_i}$, where $p_i$ are the probabilities associated with each pure state $\ket{\psi_i}$, satisfying $\sum_ip_i=1$. In this paper, the density matrix is employed to analyze the evolution of quantum states as they undergo various transformations and interactions, which are central to noise modeling and calculation in subsequent sections.

\subsection{Graph State Definition} 
A graph state $\ket{G}$ is a specific type of multipartite entanglement state that is intrinsically associated with a graph $G=(V, E)$\footnote{For clarity, the notation $G$ denotes a graph (structure) while $\ket{G}$ in dirac notation denotes a quantum graph state.} where $V$ is the set of vertices and  $E$ is the set of edges connecting these vertices \cite{hein2004multiparty},\cite{hein2006entanglement}. In graph $G$, a vertex represents a physical qubit whereas an edge indicates the entanglement between the two qubits. The formal definition of a graph state $\ket{G}$ is given by:
\begin{equation}
\ket{G} = \prod_{(a,b) \in E} CZ_{(a,b)} \ket{+}^{\otimes |V|}
\end{equation}

Here, $\ket{+}^{\otimes |V|}$ denotes the tensor product of ${|V|}$ copies of $\ket{+}$ state at all the vertices of $G$, and $CZ_{(a,b)}$ is the controlled-Z gate operation applied to the pair of qubits corresponding to the edge $(a, b) \in E$. 



\subsection{Graph State Generation}
The generation of a graph state is achieved through the following sequence of steps:
\begin{enumerate}
    \item \textit{Basis state preparation}: Initiate by preparing each qubit, which corresponds to a vertex in $V$, in the basis state $\ket{0}$. 
    \item \textit{Superposition transformation}: Apply the Hadamard gate $H$ on each qubit, which transforms $\ket{0}$ into the superposition state $\ket{+}=\frac{1}{\sqrt{2}}(\ket{0}+\ket{1})$. 
    \item \textit{Entanglement creation}: For each pair of qubits for which an entanglement is desired, apply a $CZ$ gate. Note that the $CZ$ operation is commutable, so it can be applied in any arbitrary sequence (or all at once). A rendered entanglement is represented by an edge in the graph $G$. 
\end{enumerate}



It is important to highlight that graph states with the same number of qubits and topology, differing only in qubit labeling, possess identical quantum properties. However, qubit labeling is far from trivial in engineering systems, as qubits may be located in vastly different physical devices of unique limitations and capabilities. This distinction is critical for designing engineering protocols, which are fundamental in the majority of quantum information processing and communication applications. Specifically in this study, we consider that each qubit is associated with a distinct network node. A graph state of the ensemble of qubits is thus the reflection of the entanglement state of the underlying network nodes. For this reason, we treat these graph states with different labels as distinct entities and refer to them as ``labeled graph states''. Fig. \ref{fig:framework} illustrates two such labeled graph states (the top one and the bottom one in Step 2 of Fig. 1).



\subsection{Graph State Properties}

\subsubsection{Local Complementation (LC)} 
Denoted as $LC_a$, when it is applied to a vertex $a$ in a graph $G$, it results in a graph formed by the complement of the subgraph induced by the neighbors $N_G(a)$ of $a$, while the remainder of the graph remains unaltered. Specifically, this operation involves the modification of edges among the neighbors of $a$: if an edge exists between any two vertices in $N_G(a)$, it is removed from $G$; conversely, if there is no edge between a pair of vertices in $N_G(a)$, then an edge is added. Step 2 of Fig. \ref{fig:framework} illustrates the outcomes of sequentially applying $LC$ on vertex $v_1$ and $v_3$.


In a graph state, the local complement on qubit $a$ can be implemented by the following unitary operator $U_{a}^{LC}$:
\begin{equation}
U_{a}^{LC} = \sqrt{-iX_{a}} \bigotimes_{b \in N_G(a)} \sqrt{iZ_b}
\label{eq:lc}
\end{equation}
That is, $U_{a}^{LC} \ket{G} = \ket{LC_a(G)}$, in which the right-hand side is interpreted as the LC-equivalent graph of $G$ on vertex $a$. Here, the term $i$ is a complex number which represents the phase factor.  $X_{a}$ denotes the pauli-$X$ gate applied to qubit $a$, whereas 
$\otimes_{b \in N_G(a)} Z_b $ acts as a Pauli-$Z$ gate to every qubit in the neighborhood of $a$.
The operators $\sqrt{-iX}$ and $\sqrt{iZ}$ can be calculated as
$\sqrt{-iX} = \frac{1}{\sqrt{2}} 
\begin{pmatrix} 1 &  -i \\ 
                -i & 1 
\end{pmatrix}$, and
$\sqrt{iZ} = \frac{1}{\sqrt{2}} 
\begin{pmatrix} 1+i & 0 \\ 
                0 & 1-i
\end{pmatrix}$, respectively. Therefore, executing local complementation on qubit $a$ is tantamount to applying single-qubit gate $\sqrt{-iX}$ on qubit $a$ and $\sqrt{iZ}$ on all its neighboring qubits.




Two graph states are LC-equivalent if and only if they can be transformed into another after a series of local complementations. The LC-equivalent graph states share the same quantum properties, such as the Schmidt measure value \cite{adcock2020mapping}. Identifying whether two graph states are LC-equivalent can be solved in polynomial time, but counting single-qubit LC-equivalent graph states is $\#$P-complete\cite{adcock2020mapping, dahlberg2020counting}. Therefore, searching all potential LC-equivalent graph states of the input graph state is $\#$P-complete. 
Some researchers have exhausted all LC-equivalent graph states for with a small number of qubits, typically fewer than 11, and curated them in graph state orbits \cite{adcock2020mapping}. An orbit is a graph of graphs, i.e., a collection of LC-equivalent quantum graph states. Two vertices in an orbit have an edge if they can be transformed to each other by an LC operation. The whole collection of $q$-qubit graph states can be divided into several orbits, e.g., Table \ref{table:orbit_info_table} and Fig. \ref{fig:orbits} in Appendix. A.

\subsubsection{Projective measurement}
Single-qubit projective measurement may change the entanglement of the whole graph state. Specifically, applying Y-measurement on a vertex $a$ is to delete vertex $a$ and all its incident edges, and locally complement its neighbors. Applying Z-measurement on a vertex $a$ is to delete the vertex $a$ and all its incident edges. 
To apply an X-measurement on vertex $a$, first select a neighbor vertex $b_0 \in N_G(a)$, perform an LC operation on $b_0$, apply a Y-measurement on $a$, and then repeat the LC operation on $b_0$.

%% file: system.tex
\section{Quantum System Model and Assumptions}\label{system}

In this section, we present a quantum system implementation on which our subsequent algorithm is based. 
We then discuss the introduced noise and errors by this system to the establishment of a quantum graph state. 

\subsection{Physical System Model}
To construct the target graph state, we have to first create a single point-to-point entanglement between two remote nodes in the physical network. We consider that each node follows the cavity quantum electrodynamics platform such as trapped ions and the nitrogen-vacancy centers in diamond that holds spin qubits. Spin qubits are stationary, individually controllable with ultra-low gate errors, and well-suited for storage due to their relatively long coherent time \cite{ruf2021quantum}. These cavities serve as the foundational element for spin-photon interfaces, enabling the conversion of stationary spin qubits into flying photonic qubits for quantum communications.

\begin{figure}[htbp]
\centering
\includegraphics[width=0.8\linewidth]{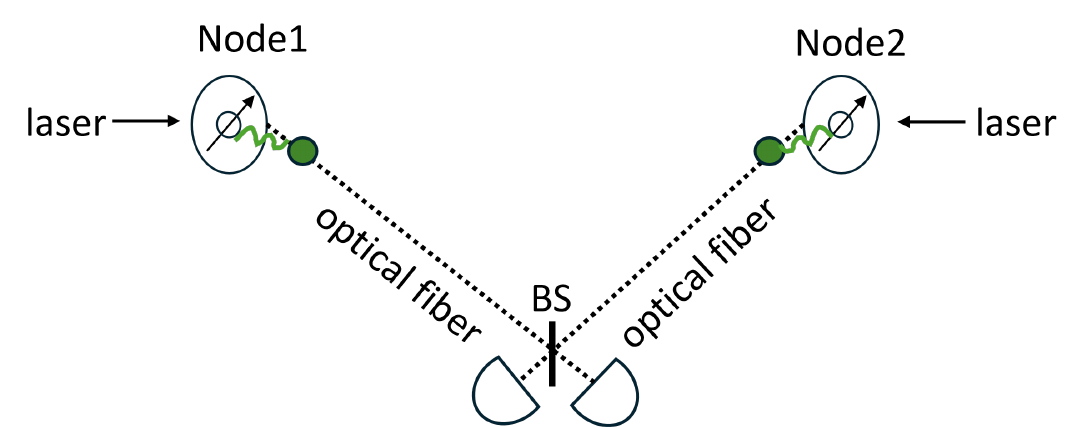}
 \caption{Spin-to-spin entanglement generation.}
\label{fig:entanglement}
\end{figure}

The entanglement of two remote spin qubits, i.e., two network nodes, is mediated and heralded by the successful transmission and detection of their associated photonic qubits. Specifically, at each node, a spin qubit with strong coupling with the cavity's photonic field can be excited by a controlled pulse laser to induce the emission of photonic qubits, effectively creating entanglement between the stationary spin qubit and the flying photonic qubit. The generated photonic qubits are then transmitted through a fiber optic channel toward a detection point. The successful joint Bell state measurement (BSM) on the photonic qubits heralds the entanglement of two spin qubits on the remote nodes. The whole procedure is shown in the Fig. \ref{fig:entanglement}. Note that in our work, entanglement between two end nodes is established by directly interfering the two flying qubits, with intermediate nodes along the quantum channel acting only as a BSM station without swapping or storage capacity.

In this process, noises and losses are mainly attributed to the following stages: (1) the coupling loss in the spin-photon transduction, (2) photon loss and noise in the optical fiber, and (3) measurement noises incurred by the BSM inefficiency (due to Hong–Ou–Mandel effect) and dark counts associated with the single-photon detector. Collectively, we model the system noise and loss as follows.


\subsection{Decoherence, Loss and Error Models}
When a flying qubit is transmitted through an optical fiber, its state is assumed to evolve into the following mixed state \cite{muralidharan2014ultrafast}, \cite{dur2005standard}:
\begin{equation*}
\rho' = \eta_1 \eta_2 (1 - \epsilon_d) \rho 
+ \frac{\eta_1 \eta_2 \epsilon_d}{4} \sum_{j=0}^{3} \sigma_j \rho \sigma_j 
+ (1 - \eta_1 \eta_2) \ket{\text{vac}} \bra{\text{vac}} 
\label{eq:error_model}
\end{equation*}
where $\rho$ represents a photon's initial state, while $\rho'$ is its state after transmission through the channel. $\eta_1$ is the spin-photon conversion probability while $\eta_2 = e^{-L_0/L_{att}}$ denotes the photon survival probability over the channel, with $L_0$ being the channel distance and $L_{att}$ the fiber attenuation, where $L_{att}=\frac{10}{\alpha ln(10)}$, $\alpha$ is the fiber loss factor. The variable $\epsilon_d$ refers to the depolarization noise during transmission. The symbols $\sigma_0$, $\sigma_1$, $\sigma_2$, and $\sigma_3$ correspond to the Pauli-X, Pauli-Y, Pauli-Z, and Identity operators, respectively. The expression $\ket{\text{vac}} \bra{\text{vac}}$ signifies the density matrix of a vacuum state, indicating that the photon is lost. 
Consequently, the success probability of generating and transmitting a single flying qubit via fiber channels is given by $\eta_1 \eta_2 (1 - \epsilon_d)$, which reflects the likelihood of the qubit being successfully generated and transmitted through the fiber channel without noise or loss. The model also incorporates error probability, which accounts for the noise and photon losses. These effects are captured in the second and third terms of the density matrix, which describe depolarization noise and photon loss, respectively.
This loss/noise model is a direct adoption from the work by Muralidharan et al. in \cite{muralidharan2014ultrafast}. An alternative approach to modeling noise and loss is the stabilizer formalism by Mor-Ruiz et al. in \cite{mor2023noisy}.

When two flying qubits impinge on the detection point for joint measurement, due to Hong-Ou-Mandel effect, BSM has a fundamental limit in resolution --- denoted as $p_\text{BSM}$, e.g., it is only 50\% if BSM is implemented by linear optics. Moreover, we consider that the single-photon detectors at the detection point have the same dark count rate of $f_\text{dc}$ Hz ($f_\text{dc} > 1$). Then, a successful measurement, thus heralding spin-to-spin entanglement, has a probability 
\begin{equation*}
p_\text{BSM-msr} = p_\text{BSM} \left(1-\frac{1}{f_\text{dc}}\right)^2 
\label{eq:measure_error}
\end{equation*}

Collectively, we can calculate the successful probability for establishing a spin-to-spin entanglement between two remote nodes $v_i$ and $v_{i'}$ as 
\begin{equation}
p_{i,i'}= \left[\eta_1 \eta_2 (1 - \epsilon_d)\right]^2 \times p_\text{BSM} \left (1-\frac{1}{f_\text{dc}}\right)^2.
\label{eq:elementary_success}
\end{equation}

In addition to the error model for photonic qubits, the spin qubits also undergo a series of local operations, including the $CZ$ operations, Y-measurement, and LC operations. They incur 2-qubit gate noise, measurement noise, and 1-qubit gate noise, respectively. First, to account for 2-qubit gate errors, we define it as follows:
\begin{equation*}
\rho^\prime= \left(1-\epsilon_{2g}\right) CZ \rho CZ^{\dagger}+ \frac{\epsilon_{2g}}{16} \sum_{j^{\prime}=0}^3 \sum_{j=0}^3 \sigma_{j^{\prime}}^{(C)} \sigma_j^{(T)} \rho \sigma_j^{(T)} \sigma_{j^{\prime}}^{(C)}
\label{eq:2gate_error}
\end{equation*}
where $\epsilon_{2g}$ denotes the error rate associated with the CZ operation while the superscript (C) and (T) refer to the control and target qubit, respectively. Likewise, we can define the 1-qubit gate error probability as $\epsilon_{1g}$ following the model below:
\begin{equation*}
\rho' = (1 - \epsilon_{1g}) \rho 
+ \frac{\epsilon_{1g}}{4} \sum_{j=0}^{3} \sigma_j \rho \sigma_j .
\label{eq:1gate_errpr}
\end{equation*}
We can further denote the Y-measurement success probability as $p_{\text{Y-msr}}$, which not only accounts for the intrinsic success rate of the Y-measurement but also includes the required correction operations (i.e., applying $\sqrt{iZ}$ or $\sqrt{-iZ}$) on the neighboring qubits, necessary for ensuring a correct graph-state transformation. It is worthwhile to mention that despite some noises can cancel each other when they appear multiple times in the process (e.g., $X \cdot X = I$), they only contribute high-order probability terms $\mathcal{O}(\epsilon^n)$ that is almost negligible. Moreover, tracking these repetitive high-order noise terms is prohibitively difficult. Therefore, for the calculation of overall success probability accounting for all the noisy steps, we only consider the first-order probability term that is the success probability of each step. Note that by disregarding high-order terms, the calculated success probability presented here serves as an approximation.

%% file: principle.tex
\section{Algorithmic Principles}\label{principle}
In this section, we begin by presenting a detailed description of the problem scenario under investigation and an overview of the proposed method. Following this, we elucidate the importance of identifying an LC-equivalent graph state by LC operations and outline the methodology of searching for a graph state that meets the specified requirements. Subsequently, we discuss the procedure for reconstructing the target graph state from its LC-equivalent state by developing compressed local operations.

\subsection{Problem Description}

Consider a physical network, denoted as $G_p=(V_p, E_p)$, consisting of $|V_p|$ nodes and $|E_p|$ links. A link $(v_i,v_{i'}) \in E_p$ indicates the presence of a fiber channel that connects nodes $v_i$ and $v_{i'}$. Each link is associated with a cost, i.e., an edge weight $\omega(v_i,v_{i'})$, to represent the physical distance $L_0$ between node $v_i$ and $v_{i'}$. A quantum application may desire a graph state $\ket{G}$ of a specific entanglement structure $G=(V, E)$ among selected nodes in the physical network $G_p$, that is to say, $V \subseteq V_p$. 
Take Step 1 in Fig. \ref{fig:framework} as an example, the set $\{ v_1, v_2, v_3, v_4\}$ selected as nodes of the target graph state in a star topology, is the subset of $\{v_1, v_2, v_3, v_4, v_5\}$.

The prevalent method to establish a graph state $\ket{G}$ involves the generation of pair-wise spin-to-spin entanglement between physical nodes that have an entanglement relationship in $\ket{G}$, followed by qubit fusion to produce $\ket{G}$. While most existing research follows the Steiner tree --- a minimum cost distribution tree --- to distribute elementary entanglement, our algorithm will follow the minimum-cost LC-equivalent state of $\ket{G}$ for the preservation of $\ket{G}$'s entanglement structure. The fusion step, on the other hand, is rather standard that involves CZ operations and Y-measurements, with further details provided in Section \ref{model:graph_state_implementation}. 
Our algorithm is then followed by LC operations that transform the minimum-cost LC-equivalent state back to the original graph state $\ket{G}$, as indicated by Step 4 in Fig. \ref{fig:framework}.



Notably, we can calculate the overall success probability by combining the rates of success for single spin-to-spin entanglement distribution, qubit fusion, and LC transformations, presented as:
\begin{equation}
   p_{\textbf{overall}} = \underbrace{\prod_{(v_i,v_{i'})\in E^*} p_{i,i'}}_{\text{\parbox{2.3cm}{\centering ENTANGLEMENT\\DISTRIBUTION}}} \times \underbrace{({p_{\text{cz}}}\cdot {p_{\text{Y-msr}}})^{m_1}}_{\text{FUSION}} \times \underbrace{{p_{1g}}^{m_2}}_{\text{LC}}.
   \label{eq:overall_prob}
\end{equation}
Eq. (\ref{eq:overall_prob}) is based on the assumption that an error at any stage in the three-step process requires the entire procedure to be restarted. In other words, our design has no checkpoints by saving and certifying quantum states in the middle. This strict assumption makes Eq. (\ref{eq:overall_prob}) a performance lower bound to any practical quantum systems. Here, $p_{i,i'}$ follows Eq. (\ref{eq:elementary_success}) that signifies the probability of successfully establishing spin-to-spin entanglement between nodes $v_i$ and $v_{i'}$, with $(v_i,v_{i'})\in E^*$ indicating an edge in the LC equivalent graph state $\ket{G^*}$. $p_{\text{cz}}=1-\epsilon_{2g}$ represents the success rate of the CZ operation, $p_{\text{Y-msr}}$ is the Y-measurement success rate, $p_{1g}=1-\epsilon_{1g}$ denotes the 1-qubit gate success rate. Specifically, in each fusion operation, a CZ operation is always accompanied by a subsequent Y-measurement. The overall fusion success probability is thus raised to a power of $m_1$ which equals $2|E^*|-|V^*|$. On the other hand, $m_2$ denotes the number of applied 1-qubit gates in the LC operations following Eq. (\ref{eq:lc}) that transform $\ket{G^*}$ back to $\ket{G}$.

Our design goal is to find the optimizer $\{\ket{G^*}, m_1, m_2\}$ that maximizes the overall success probability $p_{\textbf{overall}}$.
As we will elaborate in Section \ref{sec:lc_graph}, calculating the optimizer tuple $\{\ket{G^*}, m_1, m_2\}$ in a single attempt for Eq. (\ref{eq:overall_prob}) is a daunting task. Our solution is instead to search for the best strategy for each of the three steps. This entails finding the LC-equivalent graph state $\ket{G^*}$ with the highest entanglement generation success probability and the lowest qubit gate count ($m_1$ and $m_2$). Subsequent sections will discuss achieving the optimal $\ket{G^*}$ and minimizing qubit gate errors.

\subsection{LC-equivalent Graph State Search}\label{sec:lc_graph}

To begin with, our proposed method is to find an LC-equivalent graph state $\ket{G^*}$ with maximum success probability in establishing spin-to-spin entanglement. For a graph state with a small number of vertices $|V|$, we can enumerate all possible LC-equivalent graph states for a structured and fast search of the optimal LC-equivalent state. 
However, as the number of qubits of the graph state increases, enumerating all LC-equivalent states becomes prohibitively complex. This is because every intermediate state in the LC transformation has $|V|$ possible evolution paths to its next state, leading to a curse of dimensionality when $|V|$ increases.
Moreover, it has been rigorously proven that provided that determining the count of single-qubit LC-equivalent graph states is $\#$P-complete \cite{adcock2020mapping, dahlberg2020counting}, searching for all potential LC-equivalent graph states of a given input graph state falls into the $\#$P-complete category. Consequently, identifying the graph state $\ket{G^*}$ with the maximal spin-to-spin entanglement success probability within polynomial time is not feasible.


To search for the graph state $\ket{G^*}$ that maximizes the likelihood of success, we employ the simulated annealing (SA) algorithm \cite{bertsimas1993simulated}. SA, inspired by the physical annealing process of slow cooling to minimize energy, uses random searches to escape local optima by occasionally accepting worse solutions with a certain probability. As the ``temperature'' decreases, this probability reduces, guiding the search toward a global optimum. SA has been widely used in the quantum domain to solve various optimization problems, such as optimizing quantum circuits, preparing quantum states, and minimizing energy functions in quantum systems. For example, SA has been applied to optimize quantum circuits, reducing gate count and improving efficiency \cite{zhou2020quantum}. Additionally, Quantum Simulated Annealing (QSA), a quantum version of SA, leverages quantum techniques like quantum walks and phase estimation for more efficient optimization \cite{somma2007quantum}.

In our work, SA is applied for the first time to search for the graph state $\ket{G*}$
that maximizes the likelihood of success, providing an efficient approach to approximating graph state parameters in quantum communication networks. This method allows for quick approximation of the graph state, avoiding the time-consuming enumeration of all LC-equivalents. Algorithm 1 elaborates on the SA algorithm, providing a detailed methodology for achieving our objective.

\begin{algorithm}
\caption{Simulated Annealing Algorithm}
\begin{algorithmic}[1]
\STATE \textbf{Goal:} Minimize $-\prod_{(v_i,v_{i'})\in E^*} p_{i,i'}$  
\STATE \textbf{Input:} target graph state $\ket{G}$, initial temperature $t_0 > 0$, ending temperature $t_n$, cooling factor $\beta \in (0, 1)$
\STATE Initialize: initial graph state $\ket{G^*(t)}=\ket{G}$, $t = t_0$, vertex set $A=\emptyset$
\WHILE{$t>t_n$}
    \STATE Set $t' = \beta \cdot t$
    \STATE Pick a new graph state $\ket{G^*(t)}_{\text{new}}$ in a neighborhood of $\ket{G^*(t)}$ by applying the LC on a random vertex $a$ on $\ket{G^*(t)}$\label{alg:pick_solution}
    \STATE $ p_1 = -\prod_{(v_i,v_{i'})\in E^*(t)_\text{new}} p_{i,i'}$
    \STATE $ p_2 = -\prod_{(v_i,v_{i'})\in E^*(t)} p_{i,i'}$
    \STATE $\Delta p = p_1 - p_2$\label{alg:cal_delta}
    \STATE $\text{prob} = e^{-\Delta p / t}$\label{alg:cal_acceptance probability}
    \STATE Generate a random number $r$ from a uniform distribution in $[0, 1)$
    \IF{$\Delta p < 0$ or $r < \text{prob}$}
        \STATE $\ket{G^*(t')} = \ket{G^*(t)}_{\text{new}}$, $A=A\cup \{a\}$
    \ELSE
        \STATE $\ket{G^*(t')} = \ket{G^*(t)}$
    \ENDIF
    \STATE $t = t'$
\ENDWHILE
\STATE \textbf{Output:} the final graph state $\ket{G^*}=\ket{G^*(t)}$, vertex set $A$
\end{algorithmic}
\label{alg:SA}
\end{algorithm}


Specifically, the algorithm starts with an initial target graph state $\ket{G^*(t_0)}=\ket{G}$, setting off from a high initial temperature $t_0$. This temperature acts as a critical control parameter throughout the algorithm's execution. At every iteration, the algorithm explores a neighboring graph state $\ket{G^*(t)}_\text{new}$ by applying an LC operation on a randomly chosen vertex of the current graph state $\ket{G^*(t)}$. This process introduces a randomization, reminiscent of the classical random walk algorithm, that is aimed at broadly exploring the search space for potentially more optimal solutions.

The decision to accept the new graph state $\ket{G^*(t)}_\text{new}$ hinges on the Metropolis-Hastings criterion \cite{metropolis1953equation,hastings1970monte}, which evaluates the change in the objective function $\Delta p$, representing the difference in success probability between the new and the current graph states. Given that there exists multiple paths between nodes $v_i$ and $v_{i'}$, $p_{i,i'}$ may vary. $p_{i,i'}$ is calculated following the Dijkstra algorithm to find the optimal (i.e., maximum success probability) multi-hop path in the physical fiber network. If $\ket{G^*(t)}_\text{new}$ demonstrates an improved success probability, it is accepted unconditionally. Conversely, if the success probability decreases, the algorithm may still accept $\ket{G^*(t)}_\text{new}$ with a probability $\text{prob} = e^{-\Delta p / t}$, facilitating exploration and avoiding premature convergence to local optima. Note that upon the acceptance of the graph state $\ket{G^*(t)}_\text{new}$, we document the specific vertex $a$ into the recorded vertices set $A$ where an LC operation was applied. This documentation is crucial for facilitating the recovery of the target graph state $\ket{G}$ in subsequent steps.

As the algorithm progresses, the temperature $t$ is methodically reduced according to a linear cooling schedule defined by the cooling factor $\beta$, facilitating a gradual transition from exploration to exploitation. This cooling process continues until the temperature $t$ below the minimum temperature $t_n$. The algorithm concludes by outputting the final graph state $\ket{G^*}=\ket{G^*(t)}$, representing the solution achieved through this adaptive exploration and optimization process.

\subsection{EPR Distribution and Fusion}\label{model:graph_state_implementation}
Upon finding the LC-equivalent graph state $\ket{G^*}$ with the highest success probability, our next step is to establish the graph state $\ket{G^*}$. 
This process begins with establishing elementary EPR pairs between any two nodes in the physical network following the specified edges in the graph state $\ket{G^*}$. 

Existing works have successfully demonstrated the generation of EPR entanglement $\frac{1}{\sqrt{2}}(\ket{00} + \ket{11})$ of two spins mediated by polarization photons following the setup in Fig. \ref{fig:entanglement} {\cite{sangouard2011quantum,krutyanskiy2023telecom}. To prepare for entanglement fusion, the elementary entanglement must be rendered by a CZ operation. By observing that CZ = (I$\otimes$H)CNOT(I$\otimes$H), a half-wavelength plate applied on the mediated photons before BSM will create the elementary spin-spin entanglement $\frac{1}{\sqrt{2}}(\ket{0+} + \ket{1-})$.
Subsequently, we proceed to perform qubit fusion operations to establish multipartite entanglement networks. At a chosen node, suppose there is a set of qubits $\{a, b, c\}$ after distributing elementary entanglement. We first choose qubit $a$ as the control qubit and apply a CZ operation on $a$ and $b$, which establishes an edge between $a$ and $b$. Then apply Y-measurement on $b$ so that qubit $b$ is deleted, the edges induced by $b$ are deleted, and an edge between $a$ and $b$'s neighbor is created. Following this, we continue to apply the CZ operation on $a$ and $c$, and then apply Y-measurement on $c$ (as illustrated in Fig. \ref{fig:framework}). 


\subsection{Compressed Local Operations}
To recover the target graph state $\ket{G}$ from $\ket{G^*}$, we execute a series of local operations as specified in Eq. (\ref{eq:lc}) on $\ket{G^*}$, following the reverse sequence of $A$ identified during the simulated annealing phase. Depending on the length of the vertex sequence between $\ket{G}$ and $\ket{G^*}$, each qubit in the graph state may undergo a deep quantum circuit with many quantum gates specified in Eq. (\ref{eq:lc}), thereby accumulating a great amount of gate errors. 
To mitigate these errors and enhance the success probability of establishing the target graph state, i.e., maximize ``FUSION'', and ``LC'', we employ circuit compression, leveraging the principles of quantum gate optimization.

This optimization process often involves simplifying a quantum circuit by identifying and eliminating redundant operations. For instance, applying a Hadamard gate twice in succession to the same qubit effectively reverses the qubit back to its original state, rendering these operations reducible to an identity operator. 
In this study, we employ the Berkeley Quantum Synthesis Toolkit\footnote{Berkeley Quantum Synthesis Toolkit: https://github.com/BQSKit.} to facilitate circuit compression, aiming to minimize the circuit depth and thus the cumulative gate errors. 

%% file: evaluation.tex
\section{Performance Evaluation}\label{evaluation}
In this section, we evaluate our algorithm's performance through numerical simulations using the Graph State Orbits Dataset \cite{adcock2020mapping} and synthesized physical networks. We then draw a comparison of our technology to the state-of-the-art solutions.

\begin{figure}[htbp]
\centering
\includegraphics[width=0.8\linewidth]{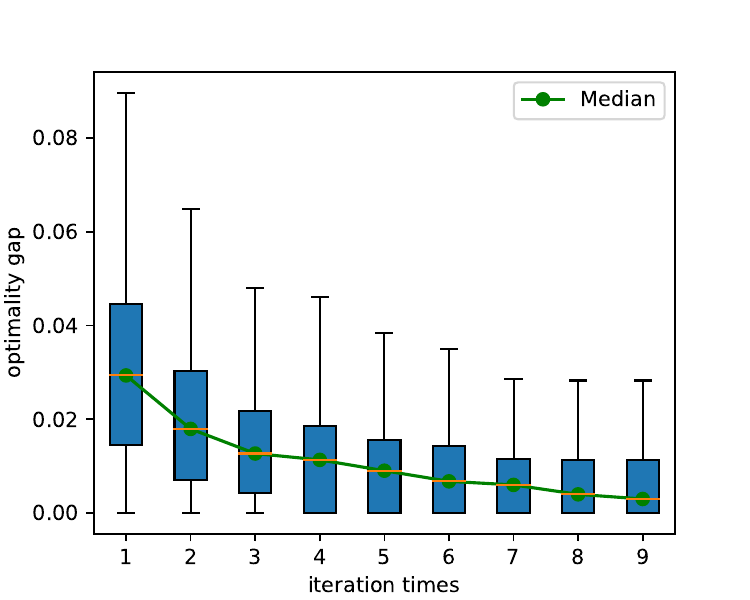}
 \caption{Convergence of SA for $q=7$ qubits on ER network. Optimality gap in y-axis indicates the difference of success probability in spin-to-spin entanglement between SA and the optimality.}
\label{fig:SA_convergence}
\end{figure}

\begin{figure*}[htbp]
  \centering
  \begin{subfigure}[b]{0.32\linewidth}
    \includegraphics[width=\linewidth]{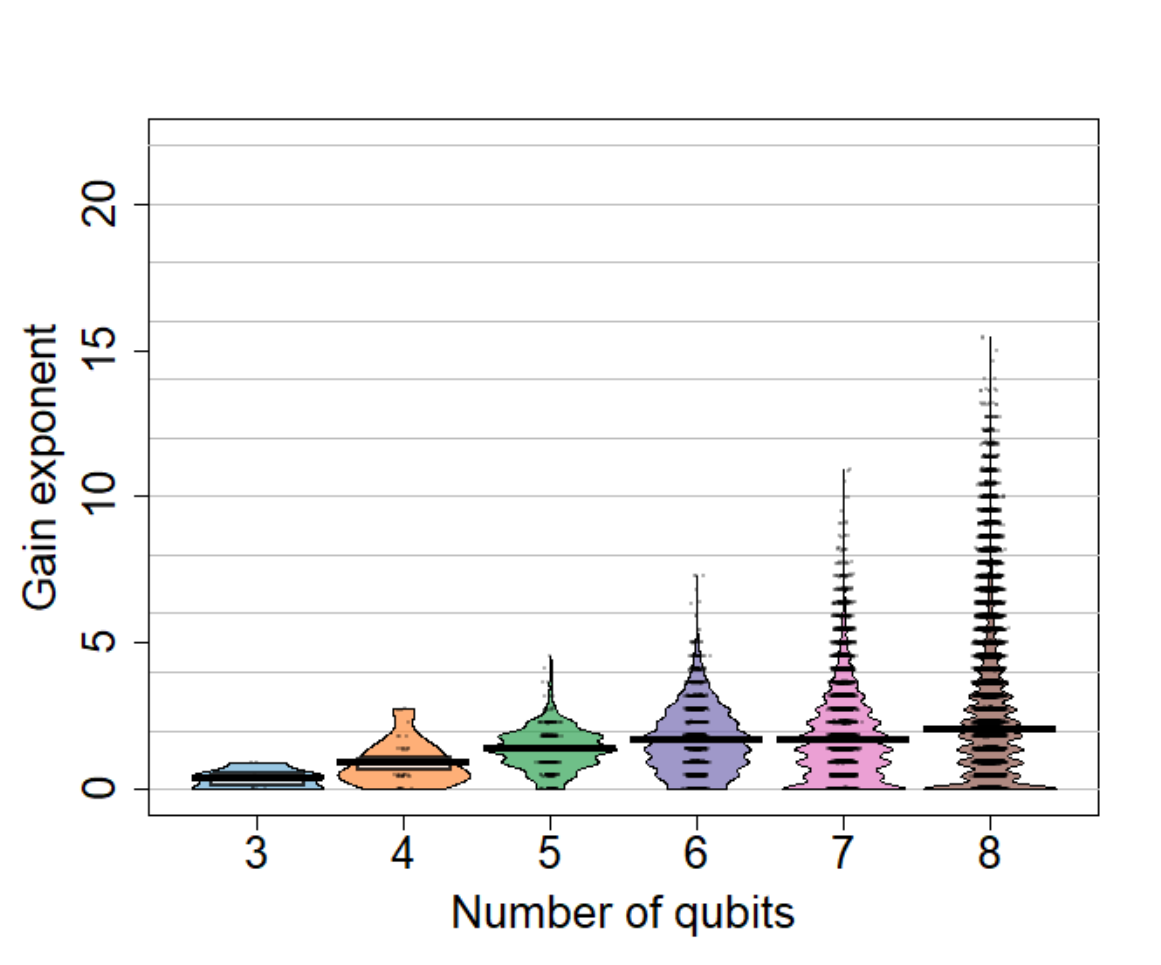}
    \caption{ER}
  \end{subfigure}
  \begin{subfigure}[b]{0.32\linewidth}
    \includegraphics[width=\linewidth]{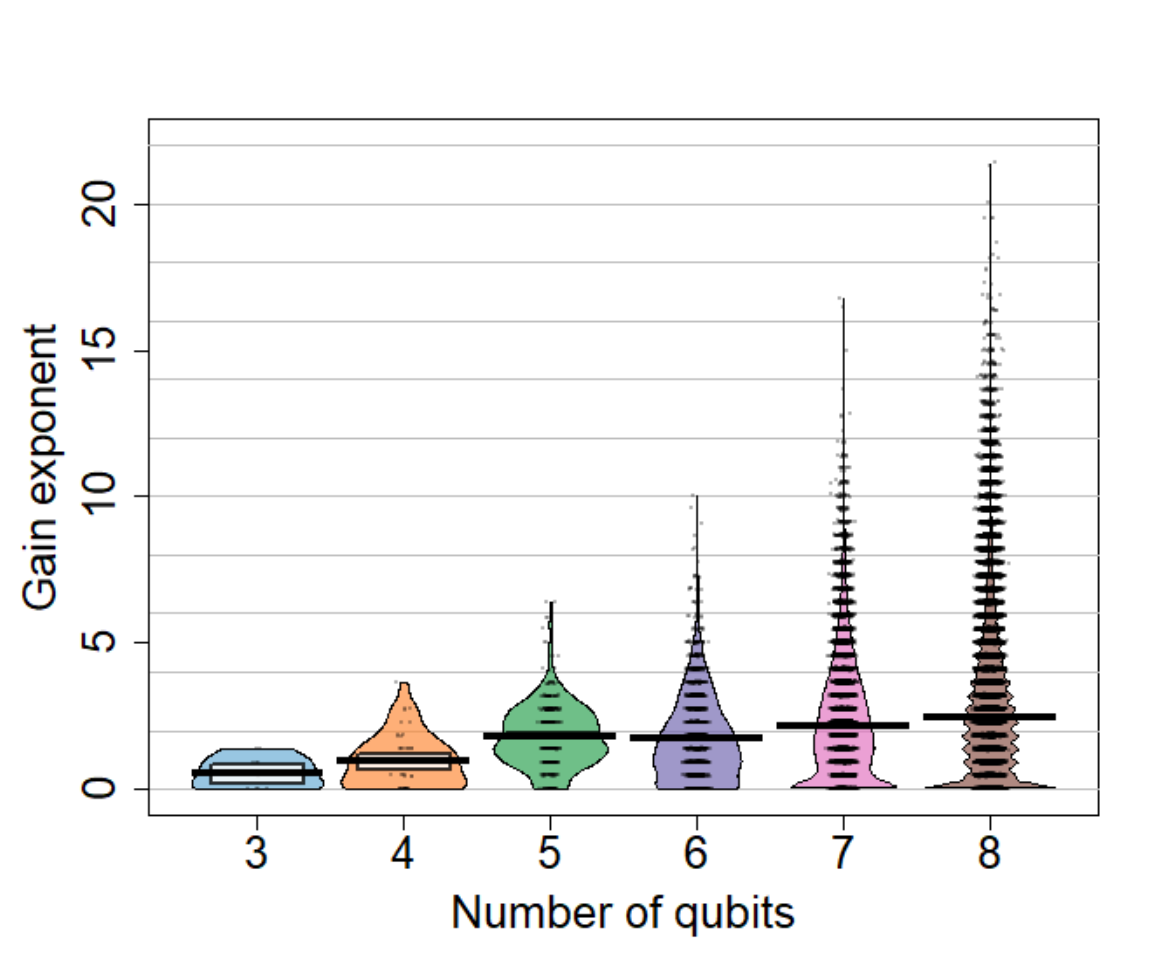}
    \caption{BA}
  \end{subfigure}
  \begin{subfigure}[b]{0.32\linewidth}
    \includegraphics[width=\linewidth]{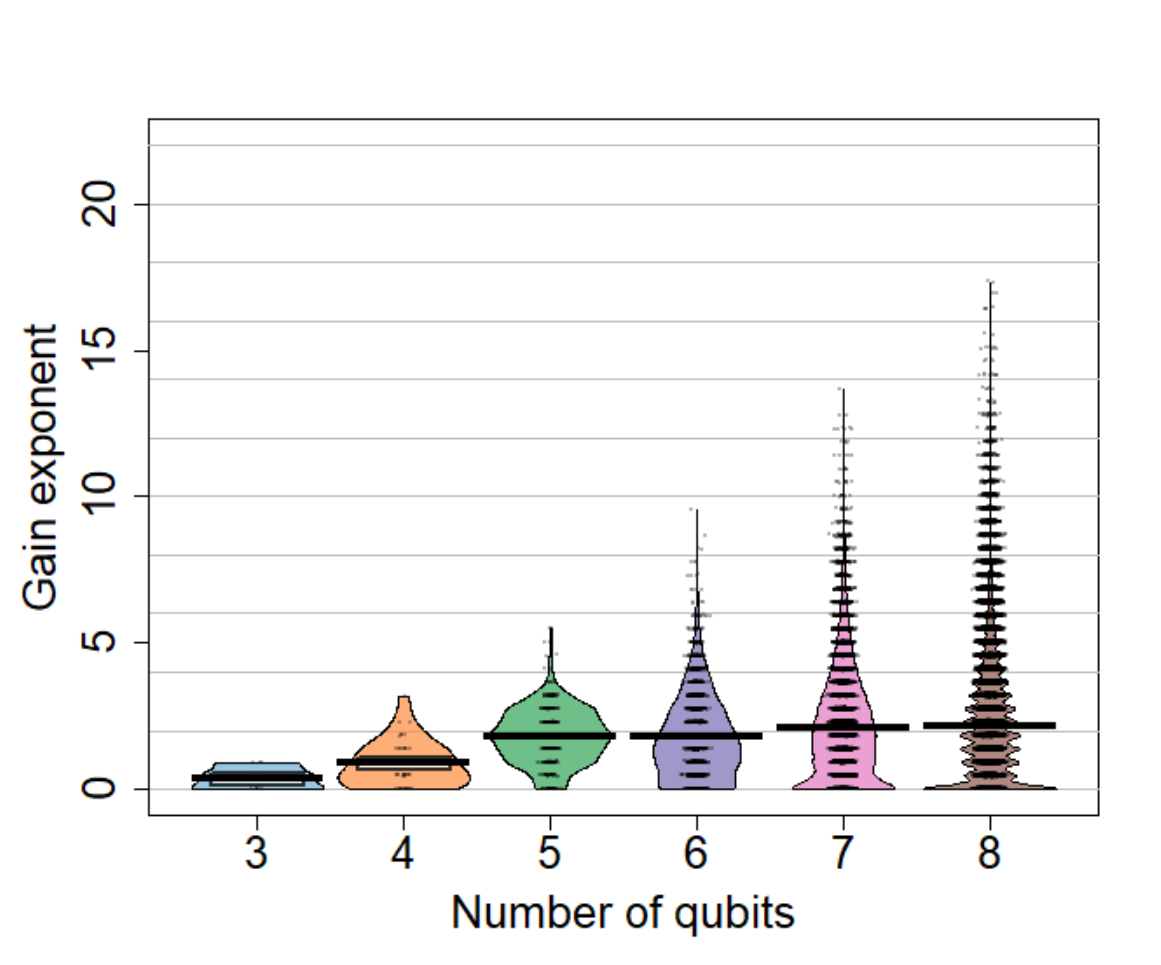}
    \caption{WS}
  \end{subfigure}
  \caption{Gain exponent of SA for three physical networks. The gain exponent ($\log_{10}$ ratio of SA success probability to direct construction) is shown for graph states with 3–8 qubits across ER (a), BA (b), and WS (c) networks.}
\label{fig:SA_gain}
\end{figure*}

\begin{figure*}[htbp]
  \centering
  \begin{subfigure}[b]{0.32\linewidth}
    \includegraphics[width=\linewidth]{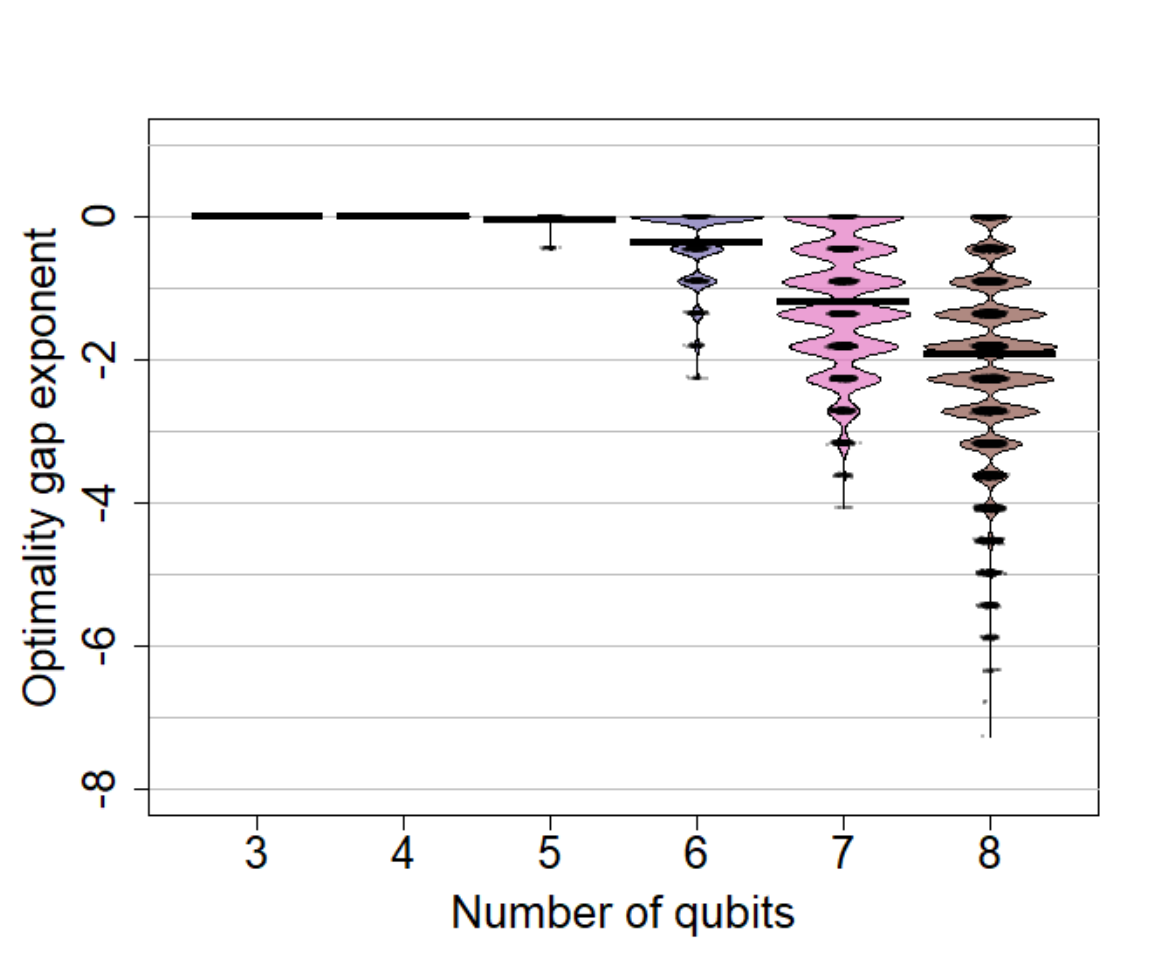}
    \caption{ER}
  \end{subfigure}
  \begin{subfigure}[b]{0.32\linewidth}
    \includegraphics[width=\linewidth]{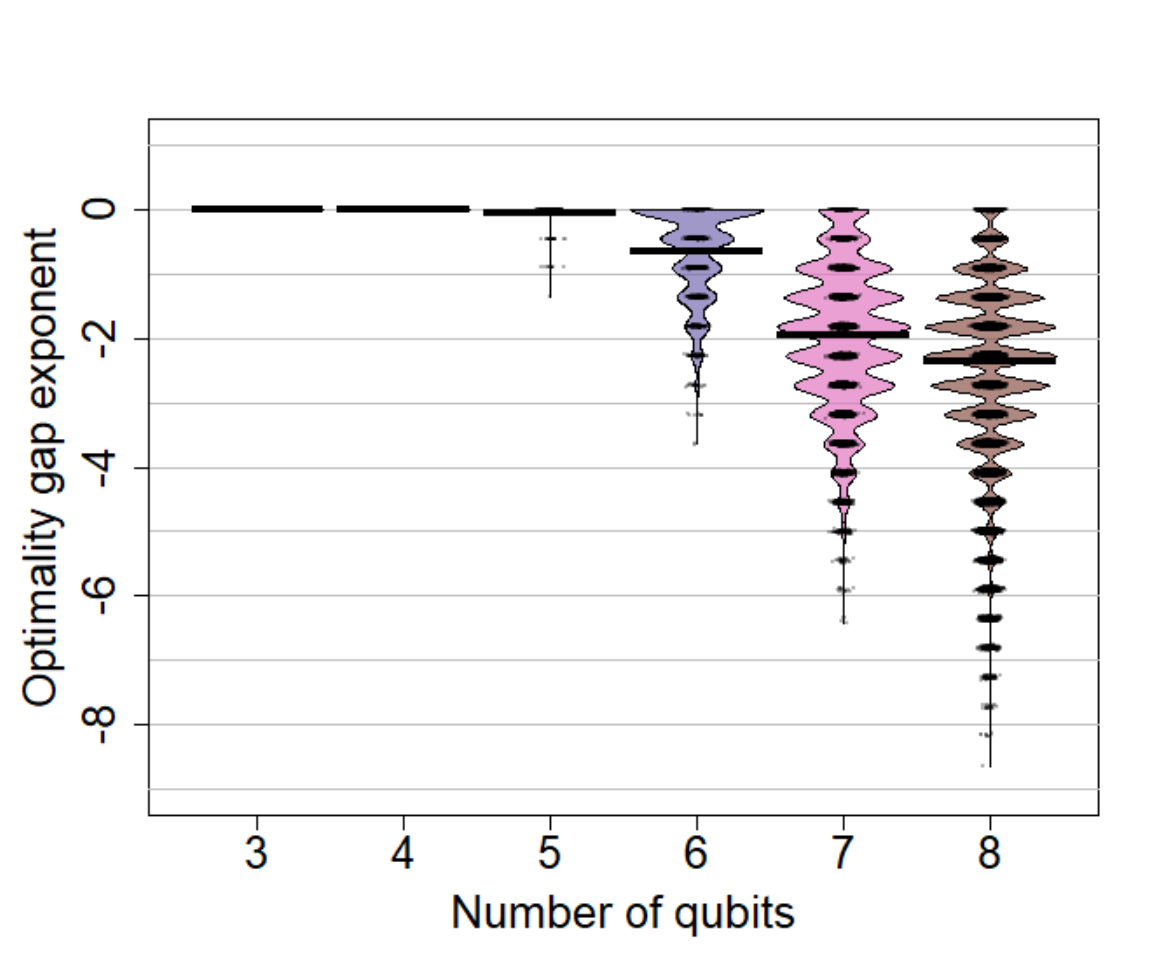}
    \caption{BA}
  \end{subfigure}
  \begin{subfigure}[b]{0.32\linewidth}
    \includegraphics[width=\linewidth]{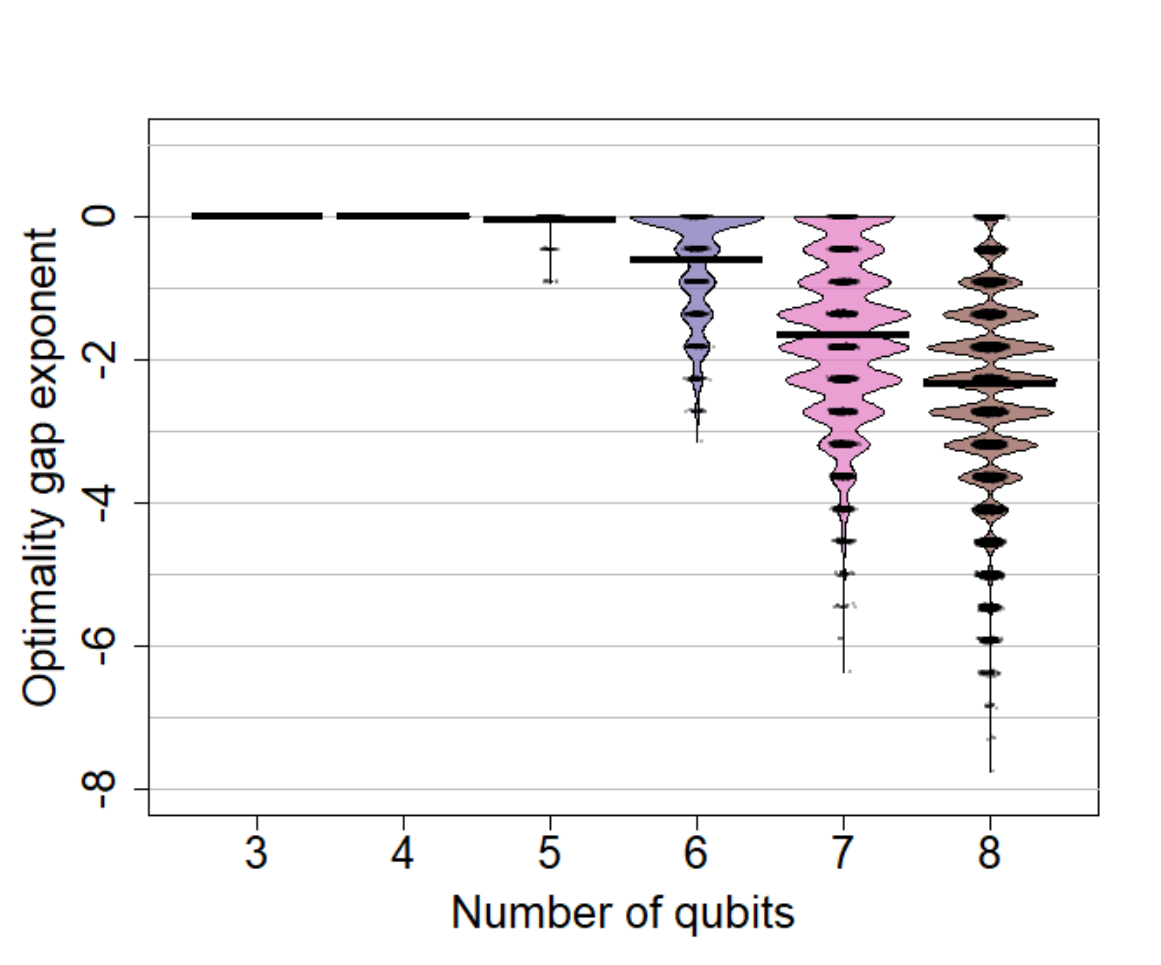}
    \caption{WS}
  \end{subfigure}
  \caption{Optimality gap exponent of SA for three physical networks. The optimality gap exponent ($\log_{10}$ ratio of SA success probability to the global optimum) is shown for graph states with 3–8 qubits across ER (a), BA (b), and WS (c) networks.}
\label{fig:SA_gap}
\end{figure*}


\begin{figure*}[htbp]
\centering
\begin{subfigure}[b]{0.32\textwidth}
    \includegraphics[width=1\textwidth]{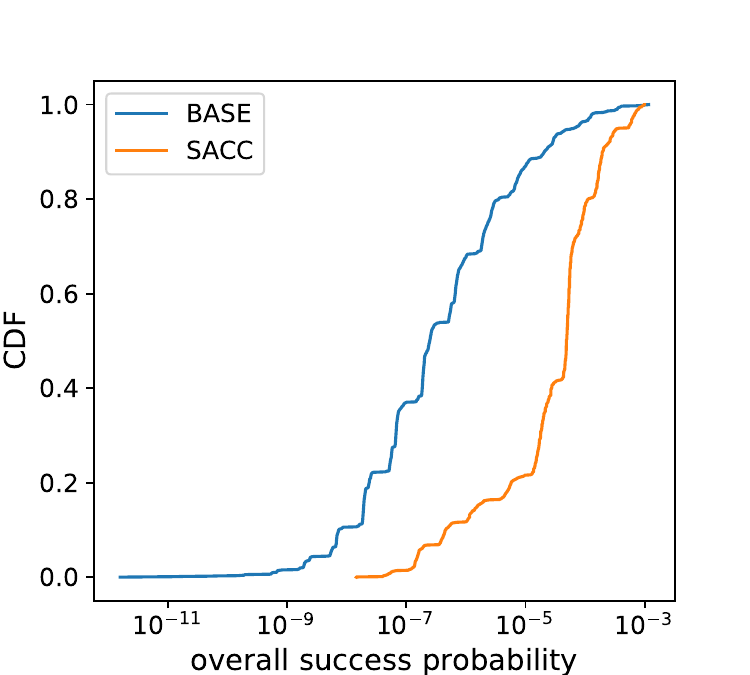}
    \caption{ER}
\end{subfigure}
\hfill
\begin{subfigure}[b]{0.32\textwidth}
    \includegraphics[width=1\textwidth]{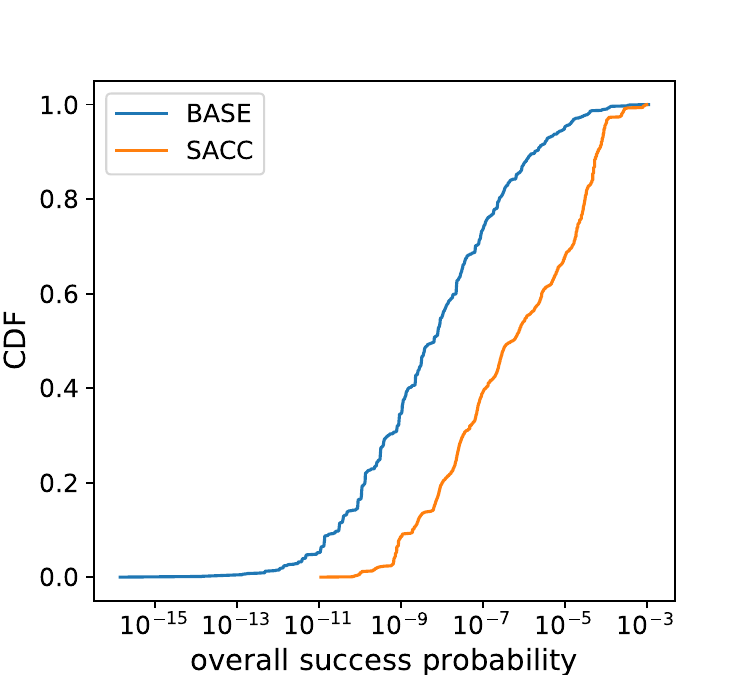}
    \caption{BA}
\end{subfigure}
\hfill
\begin{subfigure}[b]{0.32\textwidth}
    \includegraphics[width=1\textwidth]{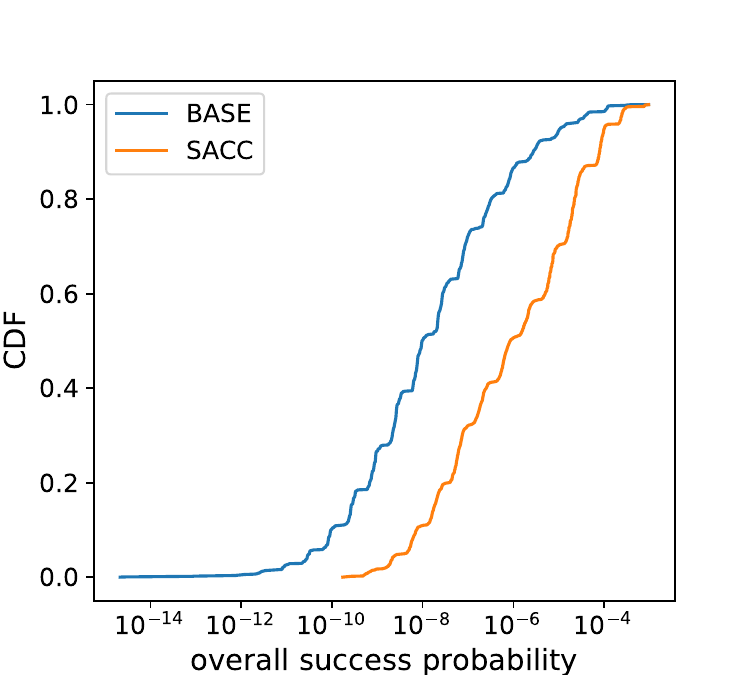}
    \caption{WS}
\end{subfigure}

\caption{CDF of overall success probability for 
$q=6$ qubit graph states using BASE (direct generation) and SACC (simulated annealing with circuit compression) across ER (a), BA (b), and WS (c) networks.}
\label{fig:overall_success_prb_comparison}
\end{figure*}

\begin{figure}[htbp]
\centering
\includegraphics[width=0.8\linewidth]{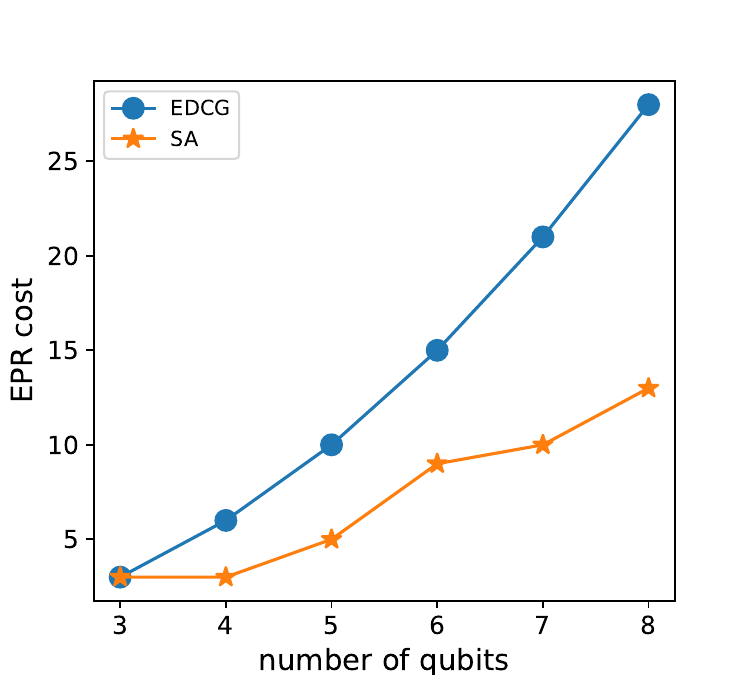}
 \caption{EPR cost comparison between SA and EDCG for 3–8 qubit graph states.}
\label{fig:EPR_compare}
\end{figure}

\textbf{Graph State Orbits Dataset:} Our evaluations and comparisons are based on the Graph State Orbits Dataset, which includes all graph states for $q = 4$ to $8$ qubits \cite{adcock2020mapping}. We also manually add orbits with $3$ qubits for completeness. An orbit is a graph of graphs, i.e., a collection of quantum graph states that are LC-equivalent. Two vertices in an orbit have an edge if they can be transformed to each other by an LC operation. The whole collection of $q$-qubit graph states can be divided into several orbits, e.g., Table \ref{table:orbit_info_table} and Fig. \ref{fig:orbits} in Appendix. \ref{app:orbit}.

\textbf{Synthesized physical network:} We use the Erd\H{o}s-R\'{e}nyi (ER), Barabási-Albert (BA), and Watts-Strogatz (WS) models to generate three synthesized network topologies to emulate physical fiber networks \cite{albert2002statistical, watts1998collective}. The generated random networks by these models bear subtle variations, allowing them to capture different network properties. The introduction of ER, BA and WS models are shown in the Appendix. B. We also give illustration examples of the three physical networks in Fig. \ref{fig:physical_networks} in Appendix. B.

\textbf{Parameter Selection:} For the three synthesized networks, the number of network nodes is set to $12$.
The fiber loss factor is $\alpha$ = 0.2dB/km, and the depolarization noise $\epsilon_d$ is isotropic, set to $0.1$\cite{muralidharan2014ultrafast}. The single fiber channel distance $L_0$ ranges from 0.5 km to 1.2 km, selected randomly according to the ER, BA, and WS model policies {\cite{muralidharan2014ultrafast}. 
The coupling efficiency $\eta_1$ between photonic and spin qubits is 90\%\cite{muralidharan2014ultrafast}, and the BSM efficiency $p_\text{BSM}$ is 50\%. The single-photon detector has a dark count rate of $f_{dc}=1,000$ Hz. Using these parameters and Eq. (\ref{eq:elementary_success}), we calculate the success probability for establishing elementary spin-to-spin entanglement. Note that the detection point is established at one of the two remote nodes, ensuring that only one flying qubit is subjected to photon loss and noise. With 2-qubit gate error (specifically, CZ gate) at isotropically 5\%, 1-qubit gate error at isotropically 1\%, and Y-measurement error rate at 1\%, we can then calculate the success probability of local operations on spin qubits \cite{muralidharan2014ultrafast}. 

\textbf{Evaluation Method:} 
For a $q$-qubit graph state $\ket{G}$, we take it as input to Algorithm \ref{alg:SA} which will result in a LC equivalent state $\ket{G^*}$. To determine if such a produced LC equivalent state is globally optimal, we find the orbit in the dataset that contains $\ket{G}$ 
and then traverse each orbit vertex to search for the optimal LC equivalent graph state, which serves as the ground truth. In our simulations, we evaluate the performance of our proposed algorithm by enumerating every possible $q$-qubit graph state ($3 \leq q \leq 8$) as the input to Algorithm \ref{alg:SA} and comparing the algorithmic output against its corresponding ground truth.

The performance evaluation metrics include the convergence behavior of SA, the performance gain or gap (in exponent) achieved by SA, the overall success probability, and the consumption of EPR pairs.




\subsection{Simulated Annealing Performance}
As shown in Fig. \ref{fig:SA_convergence}, the SA algorithm is run independently $l$ times for each 7-qubit orbit to identify the best LC-equivalent graph state. Using an ER network with 12 nodes, we select 7 nodes to generate all 7-qubit orbits and define the corresponding graph states as targets. SA is applied to each target to obtain its LC-equivalent state, and the optimality gap is computed against the theoretical optimum. Convergence is consistently achieved at $l=5$ for $q=7$, and simulations across $3 \leq q \leq 8$ show a linear relation between $l$ and $q$. Thus, we use $l=5$ in all subsequent experiments to optimize the success probability $\prod_{(v_i,v_{i'})\in E^*} p_{i,i'}$.

Before proceeding to the evaluation of SA in terms of performance gain and optimality gap, we first provide some preliminaries of the pirate plots in Fig. \ref{fig:SA_gain} and Fig. \ref{fig:SA_gap}. Pirate plots display raw data points as small dots to highlight data spread and density. The box plot (the black horizontal bar) component depicts the mean value, while the data density is visualized through a shaded area (bean plot) that varies in width vertically, appearing ``shorter and fatter'' when the distribution is more concentrated.

Fig. \ref{fig:SA_gain} presents the performance gains of the SA algorithm across various sizes of graph states within ER, BA, and WS networks. 
Here, the performance gain exponent is defined as the log10 ratio of the success probability achieved by SA to that of directly constructing the target graph state.
As shown in Figs. \ref{fig:SA_gain}, all results are positive, indicating that SA consistently yields higher success probabilities for spin-to-spin entanglement compared to directly creating the target graph state. Moreover, for a small number of qubits, such as $q=3$, the bean plots are relatively short and wide, indicating a high concentration of gain exponents ranging from approximately $0$ to $2$. 
As the number of qubits increases, the plots become more spread and thinner, with exponents reaching up to 21. This reflects the lower success probability for larger graph states but more LC-equivalent states, giving SA a broader search space for greater gains.
Additionally, the similar shapes of the bean plots across the three networks suggest that the SA algorithm performs consistently regardless of the network type, making it versatile for any physical network.

Figs. \ref{fig:SA_gap} illustrates the optimality gap achieved by SA algorithms for different graph state sizes in ER, BA, and WS networks. The optimality gap exponent is defined as the log10 ratio of SA's success probability to the ground truth success probability, i.e., $\text{optimality gap exponent}=\log_{10}{\text{gap}}$. 
For qubits less than 6, i.e., $q=3,4,5$ the pirate plot is stacked at 0, indicating no optimality gap, meaning SA finds the globally optimal LC-equivalent state. In other words, it suggests that SA is able to identify the globally optimal LC-equivalent graph state for any input graph state. Nonetheless, as the number of qubits increases, the bean plots become spread in a wider range, showing the growing scale of optimality gaps. For instance, for $q=8$, the average gap reaches an exponent of 2, with a worst-case gap of 8. Furthermore, we observe that the pirate plot is segmented by several discrete beans, clustered by many data points. This observed plot pattern suggests that the SA gets trapped in local optima within the large search space, highlighting a limitation and the potential for future improvement.

\subsection{Circuit Compression Performance}
While SA searches for an optimal LC-equivalent state, we record the path of traversed qubit vertices. Our simulations show that SA typically requires 76–78 LC operations, while ground truth only needs 3-4 LC operations, emphasizing the need for circuit compression to reduce depth from redundant LC steps.


Fig. \ref{fig:overall_success_prb_comparison} compares the overall success probability of two strategies for generating 6-qubit graph states—direct generation (\textbf{BASE}) and simulated annealing with circuit compression (\textbf{SACC})—across various types of physical networks. The result shows significant gains from circuit compression. We use the cumulative distribution function (CDF) to assess the performance of two strategies for generating graph states. For each strategy, we calculate the $p_{\text{overall}}$ in Eq.(\ref{eq:overall_prob}) for all graph states of $q=6$ qubits and sort them in ascending order.  Each curve shows the cumulative fraction of graph states (y-axis) with $p_{\text{overall}}$ (x-axis) below a given value. As shown in Fig. \ref{fig:overall_success_prb_comparison}, across ER, BA, and WS networks, the \textbf{SACC} curves (orange) consistently appear to the right of the \textbf{BASE} curves (blue), indicating significantly higher overall success probabilities. \textbf{SACC} achieves up to 5 orders of magnitude improvement over \textbf{BASE}, largely due to circuit compression. 

Additionally, the \textbf{BASE} strategy shows much higher variance in success probability than \textbf{SACC}. In the BA network, for example, \textbf{BASE} spans from $10^{-17}$ to $10^{-3}$, while \textbf{SACC} ranges from $10^{-11}$ to $10^{-3}$—a four-order-of-magnitude reduction in spread. This indicates that direct generation often produces outliers with extremely low success probabilities, whereas \textbf{SACC} significantly reduces such cases and improves overall reliability.



\subsection{EPR Consumption Comparison}
We assess the efficiency of the proposed SA strategy in minimizing EPR pair consumption by comparing it with the EDCG method from \cite{meignant2019distributing}.
For a fair comparison, we adopt the assumptions in \cite{meignant2019distributing}, including ideal EPR distribution at synchronized intervals and noiseless local qubit operations.

The EDCG method initiates by identifying a Steiner tree connecting the required qubits, followed by star expansion (SE) to create the GHZ graph state and then qubit fusion.
The number of EPR pairs needed for SE equals the number of edges in the Steiner tree. 
Collectively, the cumulative EPR cost for creating an EDCG cluster can be derived from the edges of all Steiner trees identified. In contrast, the EPR requirement in our method is confined to the edges of the LC-equivalent graph state. For a fair comparison, we align the target graph state size with the physical network, so for a network with $N$ nodes, the EPR cost for EDCG is $\frac{N(N-1)}{2}$. In our method, we consider the maximum EPR consumption for generating all possible graph states of $q$ qubits. As shown in Fig. \ref{fig:EPR_compare}, our method consistently requires fewer EPR pairs than EDCG across all $q$ values, with up to a $53.57\%$ reduction in EPR consumption for EDCG graph states.

%% file: conclusion.tex
\section{Discussion}\label{discussion}
The above simulation findings offer several insights into the SA algorithm. To begin with, SA is effective in identifying the globally optimal LC-equivalent graph state when the number of qubits is small, thanks to the moderate size of the search space. However, as the number of qubits increases, SA could get trapped in local optima, a situation exacerbated by the expanded search space. Despite falling short in finding the global optimality, SA still manages to find an LC-equivalent graph state with a higher success probability than the initial target graph state, as evidenced by the success probability distributions presented in Fig. \ref{fig:SA_gain}.

Next, the path that SA follows toward the final graph state includes many redundant steps. This redundancy results in a great number of single-qubit Clifford gates in LC operations, thereby compromising the overall success probability. Our solution in this paper is to implement circuit compression to consolidate single-qubit Clifford gates, while another approach is to adjust the SA algorithm by avoiding the vertices that have been visited. In other words, for the fundamental principle of SA is random walk, 
refining the random walk policy during the search process can be beneficial. For example, one can explore other random walk-based algorithms such as genetic algorithms, ant colony optimization, and particle swarm optimization \cite{katoch2021review,birattari2007invariance, bonyadi2017particle}. These methods could potentially improve the efficiency and effectiveness of the search process, leading to more optimal solutions.

Moreover, the SA algorithm in this paper follows the objective to maximize the success probability $\prod_{(v_i,v_{i'})\in E^*} p_{i,i'}$ in Eq. (\ref{eq:overall_prob}). However, this objective can also be broadened to optimize various other system merits. For instance, the SA algorithm can be used to search for an LC-equivalent state with the fewest edges, effectively reducing the number of $CZ$ gates required. This design is particularly significant in generating photonic graph states, where implementing $CZ$ gates presents great challenges. Furthermore, in this context, the SA algorithm has the potential to enhance the findings of Ghanbari et al. in their work \cite{ghanbari2024optimization}.


\section{Conclusion and Future Work}\label{conclusion}
In this paper, we introduced a novel two-step optimization process aimed at creating any arbitrary quantum graph states with the maximum overall success probability in a lossy and noisy quantum network. We employed the SA algorithm to identify an LC-equivalent graph state that has a higher success probability in creation compared with directly generating the original graph state.
We further applied the circuit compression technique to mitigate the errors introduced by redundant LC operations by consolidating single-qubit Clifford gates and reducing the circuit depth. Through simulations on ER, BA, and WS networks with 6–8 qubits, our approach achieved an average overall success probability of $10^{-4}$—improving upon the direct generation baseline by up to 4–5 orders of magnitude. Moreover, the proposed method reduced the EPR pair consumption by up to $53.57\%$ compared with the state of the art.


Looking ahead, future work will focus on the development of a real-time graph state distribution network, where system events such as photonic EPR pair distribution are time-labeled. This model will better capture practical characteristics of quantum memory systems, such as decoherence, within the context of cavity QED platforms. Moreover, while our method leverages simulated annealing and circuit compression, other optimization techniques, such as reinforcement learning or hybrid classical-quantum algorithms, could offer further improvements in both success probability and efficiency. These approaches, which we did not explore in this study, may provide complementary advantages and warrant further investigation.

%% file: app.tex
\appendix

\section*{A. Overview of Graph State Orbits Dataset}\label{app:orbit}

\begin{figure}[htbp]
  \centering
  \begin{subfigure}[b]{\linewidth}
  \centering
    \includegraphics[width=0.72\linewidth]{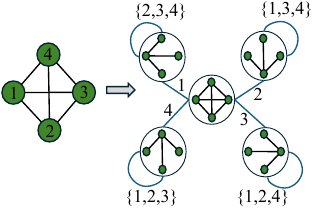}
    \caption{orbit1}
  \end{subfigure}
\hfill 
  \begin{subfigure}[b]{\linewidth}
  \centering
    \includegraphics[width=0.95\linewidth]{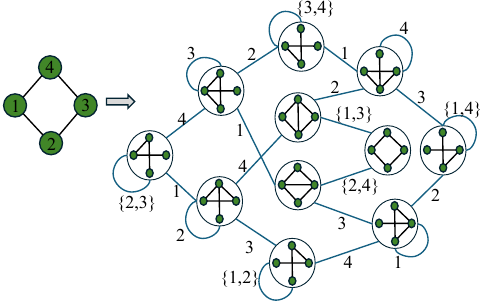}
    \caption{orbit2}
  \end{subfigure}
  \caption{All 4-qubit graph states can be grouped into two orbits. On the left side of subfigures (a) and (b), the solid circles with indices represent labeled qubits. While on the right side of subfigures (a) and (b), the diagrams show the orbits of the LC-equivalent graph states. Specifically, each small element within the black hollow circles represents a specific graph state with qubits labeled as in the left subfigure. Blue arcs between hollow circles indicate transformations between LC-equivalent graph states, and the numbers on these arcs specify the label(s) of qubits involved in the LC operation.}
\label{fig:orbits}
\end{figure}

\begin{table}[htbp]
  \centering
  \caption{Overview of the Graph State Orbit Dataset for $q=3$ to 8 qubits.}
\begin{tabular}{c|c|c|c|c|c|c}
    \hline
    q & $3$ & $4$ & $5$ & $6$ & $7$ & $8$ \\
    \hline
    \# of orbits & $1$ & $2$ & $4$ & $11$ & $26$ & $101$ \\
    Minimum \# of orbit vertices & $4$ & $5$ & $6$ & $7$ & $8$ & $9$ \\
    Maximum \# of orbit vertices & $4$ & $11$ & $132$ & $372$ & $1096$ & 3248 \\
    \hline
  \end{tabular}
  \label{table:orbit_info_table}
\end{table}
Table \ref{table:orbit_info_table} and Fig. \ref{fig:orbits} provide an overview of the Graph State Orbit Dataset. Table \ref{table:orbit_info_table} presents the number of orbits and the range of orbit vertices for each value of $q$ in the dataset. Fig. \ref{fig:orbits} gives an example of 4-qubit graph states, which are grouped into two distinct orbits.
This dataset includes both labeled and unlabeled graph states, in which the labeled one concerns about the index of a qubit while the unlabeled one does not. For instance, the GHZ states in a star topology in Fig. \ref{fig:orbits}(a) are considered different for they have distinct root qubits. For this reason, the orbit size for the labeled graph states is much larger. In this paper, we utilize labeled statesto evaluate the performance of the proposed algorithm.

\begin{figure}[htbp]
  \centering
  \begin{subfigure}[b]{0.48\linewidth}
    \includegraphics[width=\linewidth]{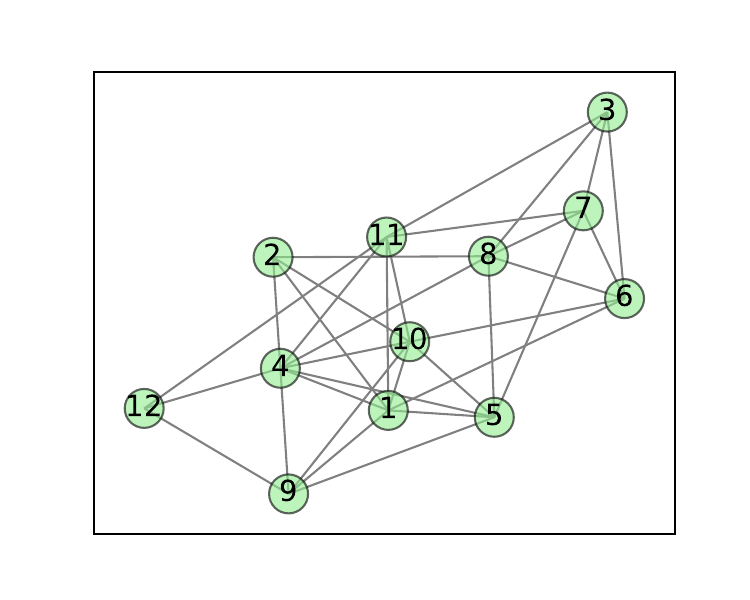}
    \caption{ER network}\label{orbit1}
  \end{subfigure}
  \begin{subfigure}[b]{0.48\linewidth}
    \includegraphics[width=\linewidth]{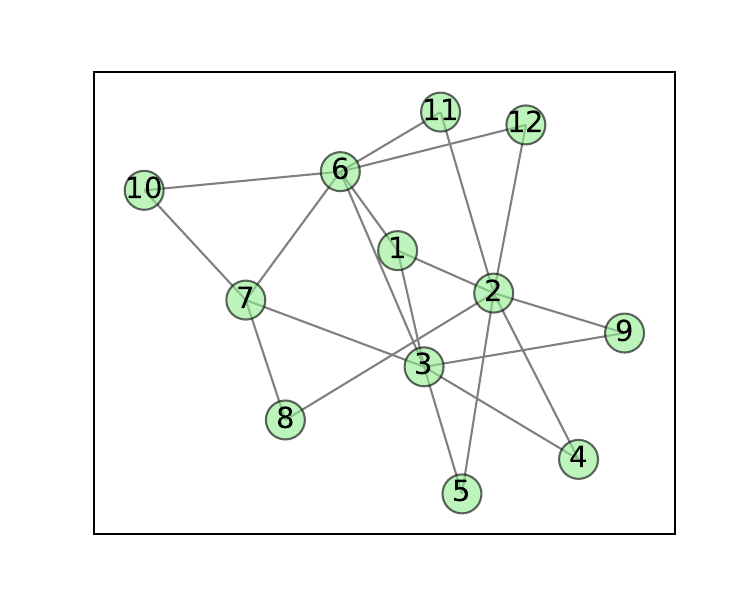}
    \caption{BA network}\label{orbit2}
  \end{subfigure}
  \begin{subfigure}[b]{0.5\linewidth}
    \includegraphics[width=\linewidth]{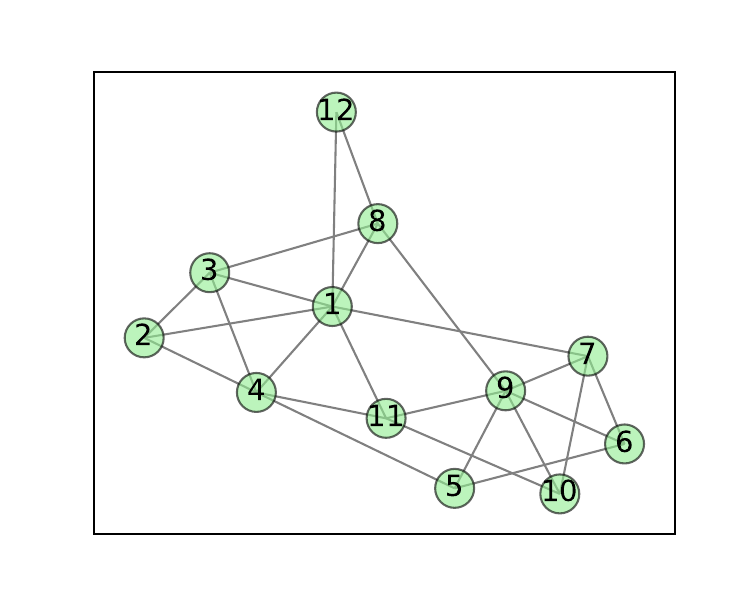}
    \caption{WS network}\label{orbit2}
  \end{subfigure}
  \caption{ER, BA, and WS networks with 12 network nodes.}
\label{fig:physical_networks}
\end{figure}

\section*{B. Overview of the synthetic physical networks}\label{app:network_model}
We use ER, BA, and WS models to generate three synthesized network topologies to emulate physical fiber networks. Specifically, the ER model generates a pure random network. The BA model produces a random and scale-free network in such a way that a node with a higher degree is more likely to receive new links. The Internet, citation networks and other human-made networks fit well into the BA model. The WS model produces random networks of small-world properties, i.e., nodes tend to be locally clustered. 
Fig. \ref{fig:physical_networks} shows three networks generated by the ER, BA, and WS models. There are 12 vertices for each synthetic network. The length of an edge is scaled to the physical distance between network nodes.
